\documentclass[aps,twocolumn,secnumarabic,amssymb, nobibnotes, prb,10pt,superscriptaddress]{revtex4}
\usepackage{graphicx}

\newcommand{\unit}[1]{\ensuremath{\, \mathrm{#1}}}

\begin{document}

%\title{Control of ultrafast exciton decay channels via Potassium intercalation into the endohedral metallofullerene semiconductor Sc$_3$N@C$_{80}$}
\title{Ultrafast charge carrier separation in Potassium-intercalated endohedral metallofullerene Sc$_3$N@C$_{80}$ thin films}

\author{Sebastian Emmerich}
\email{emmerich@physik.uni-kl.de}
\affiliation{University of Kaiserslautern and Research Center OPTIMAS, Erwin-Schr\"odinger-Stra\ss{}e 46, 67663 Kaiserslautern, Germany}
\author{Sebastian Hedwig}
\affiliation{University of Kaiserslautern and Research Center OPTIMAS, Erwin-Schr\"odinger-Stra\ss{}e 46, 67663 Kaiserslautern, Germany}
\author{Mirko Cinchetti}
\affiliation{Experimentelle Physik VI, Technische Universit\"at Dortmund, 44221 Dortmund, Germany}
\author{Benjamin Stadtm\"uller}
\affiliation{University of Kaiserslautern and Research Center OPTIMAS, Erwin-Schr\"odinger-Stra\ss{}e 46, 67663 Kaiserslautern, Germany}
\author{Martin Aeschlimann}
\affiliation{University of Kaiserslautern and Research Center OPTIMAS, Erwin-Schr\"odinger-Stra\ss{}e 46, 67663 Kaiserslautern, Germany}
%\date{February 2020}%

%\tableofcontents

\begin{abstract}
Molecular materials have emerged as highly tunable materials for photovoltaic and light-harvesting applications. The most severe challenge of this class of materials is the trapping of charge carriers in bound electron-hole pairs, which severely limits the free charge carrier generation. Here, we demonstrate a significant modification of the exciton dynamics of thin films of endohedral metallofullerene complexes upon alkali metal intercalation. For the exemplary case of Sc$_3$N@C$_{80}$ thin films, we show that potassium intercalation results in an additional relaxation channel for the optically excited charge-transfer excitons that prevents the trapping of excitons in a long-lived Frenkel exciton-like state. Instead, K intercalation leads to an ultrafast exciton dissociation coinciding most likely with the generation of free charge carriers. In this way, we propose alkali metal doping of molecular films as a novel approach to enhance the light to-charge carrier conversion efficiency in photovoltaic materials.
%Molecular thin films have promising features, beneficial for optoelectronic applications. As the films mostly maintain the intrinsic properties of the single molecules, thin films can easily be tailored to the specific application by the choice of the molecule and its functionalization. Optical properties of organic semiconductors are dominated by excitons, which prevent charge separation. Here, we aim to manipulate the exciton dynamics and decay process by intercalation, searching for an efficient charge separation, which is required e.g.~for photovoltaics or light harvesting. By tracing the optically triggered formation of charge-transfer excitons and their subsequent decay on a femtosecond time scale of a prototypical endohedral metallofullerene Sc$_3$N@C$_{80}$ thin film, we report the quenching of the exciton decay channel leading to a spatially confined Frenkel exciton upon K intercalation. A quenching of this self-trapping decay channel of these excitonic states can lead to a direct charge carrier separation, potentially rising the efficiency of the next generation of organic photovoltaic devices.
\end{abstract}

\maketitle
\section{Introduction}
Designing and functionalizing materials for dedicated areas of application is one of the biggest promises of materials science, aiming to advance the performance of next generation technology. One of the most important quests in this field is to uncover and optimize materials for light harvesting and photovoltaic applications to create renewable sources \cite{Singh2013}, which are able to convert solar energy into charge and spin carriers with utmost energy efficiency.

From the manifold of photovoltaic materials \cite{Yoshikawa2017,Yang2015,Meng2018,Albero2015}, one of the most promising candidates to supplement or even replace today's compounds are molecular materials. Besides low cost production and sustainability, the most intriguing property of these materials is their chemical tunability. This allows one to actively control the two most crucial material properties for the device efficiency on a single molecular scale, namely the spectral light absorption efficiency and the charge carrier transport \cite{Shockley1961}.
For instance, they can be manipulated or controlled by the choice of the molecular core and the molecular ligands \cite{Ross2009,Schwarze2016,Pinzon2008,Ostroverkhova2016}.

A very special and extremely tunable case in this context are fullerenes such as the buckminsterfullerene C$_{60}$. Their properties can be tailored either by changing the number of atoms of the carbon cage itself or by incorporating atoms or small clusters of atoms into the carbon cage, forming so called endohedral metallofullerenes (EMFs) \cite{Stevenson1999a,Popov2013}. The latter offers the particular opportunity to embed spin-carrying atoms into the carbon cage to form single molecular magnets, or to manipulate the charge transport properties \cite{Sato2012,Ross2009} and excited state dynamics \cite{Wu2015} of EMF films by the appropriate choice of the metallic cluster inside the carbon cage.

%Tailoring a material's properties to match the specific application is a big promise of materials science. In photovoltaics, being one of the most prominent examples of energy harvesting from renewable sources \cite{Singh2013}, there are several competing material classes with promising properties, having the potential to replace the devices currently in use, in terms of efficiency or sustainability \cite{Yoshikawa2017,Yang2015,Meng2018,Albero2015}. In the development of a light-harvesting device, besides the production costs and this sustainability, the control of both light absorbance and transport of the created charge carriers is of utmost importance, being the responsible parameters governing the efficiency of the device \cite{Shockley1961}.
%In organic thin films for instance, these properties can be accessed and tuned by choice of the molecule type and the ligands \cite{Ross2009,Schwarze2016,Pinzon2008,Ostroverkhova2016}.
%For the special case of fullerenes, species can also be incorporated in the carbon cage, forming the molecule class of the endohedral metallofullerenes (EMFs) \cite{Stevenson1999a,Popov2013}.
%Different studies on these EMFs have shown that the choice of the metal atom in the metal cluster inside the fullerene cage strongly influences the charge transport properties \cite{Sato2012,Ross2009} and that the cage symmetry as well plays a decisive role for the available relaxation channels for excited electronic states \cite{Wu2015}.

In this Letter, we explore the ultrafast dynamics of functionalized EMF materials after optical excitation with fs-light pulses. The excited state dynamics provide a direct view onto the microscopic mechanism mediating or limiting the light-to-charge conversion efficiency in molecular materials. In contrast to inorganic metals or semiconductors, the optical excitation of molecular materials with visible light does not result in the formation of free carriers, but in bound electron-hole pairs, so called excitons, with binding energies up to several hundred meV \cite{Bardeen2014}. These excitons can exhibit different charge distributions and different degrees of spatial delocalization, which are typically described in the limit of charge-transfer (CT) and Frenkel excitons and can have long depopulation (decay) times up to several hundreds of microseconds. The long depopulation times of the energetically lowest excitonic level as well as its large exciton binding energy severely limit the optically induced free carrier generation in molecular materials, a challenge that must be tackled to improve the performance of molecular-based photovoltaic applications.

\begin{figure*}[th]
  \includegraphics[width=110mm]{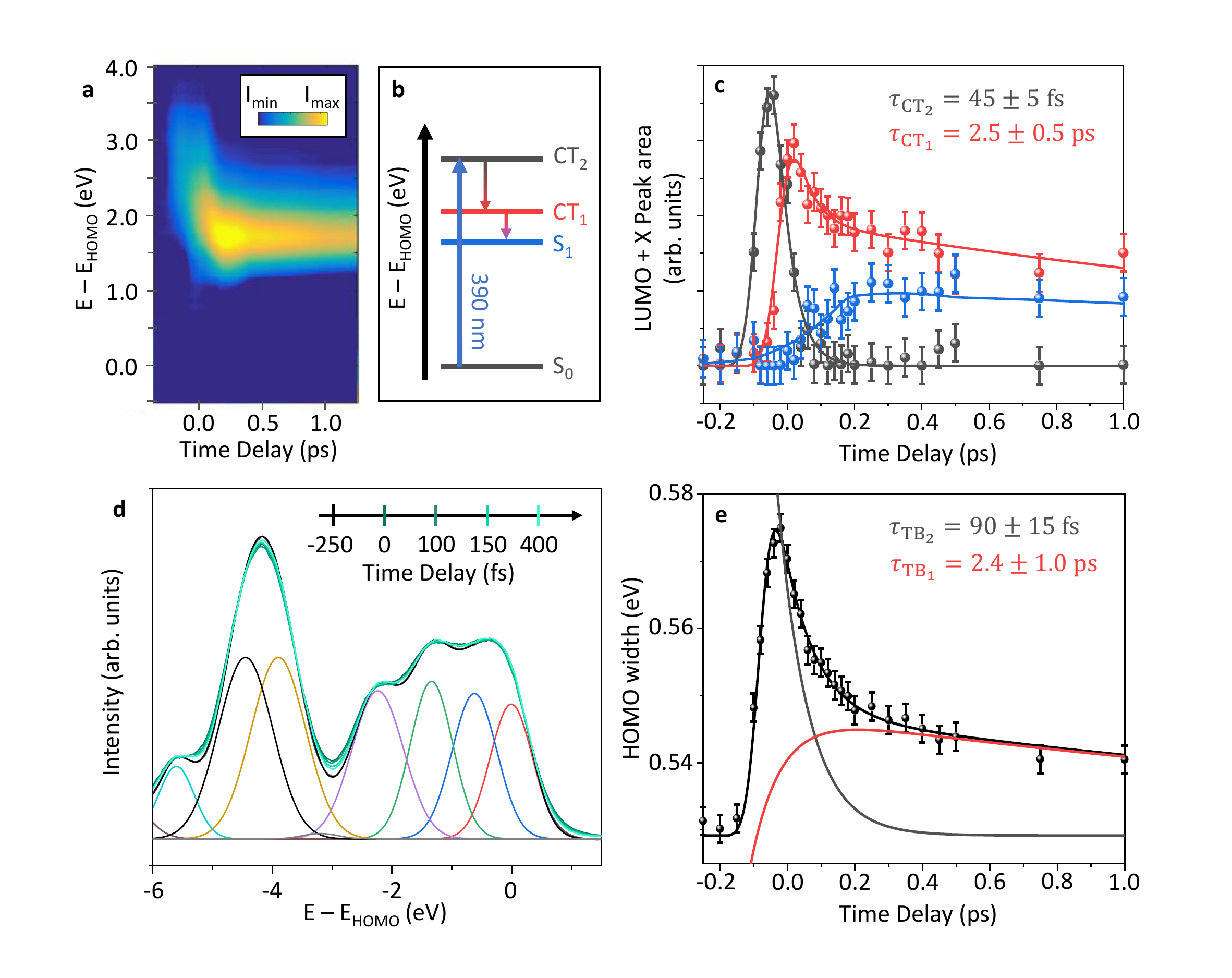}
  \caption{\textit{Electron dynamics in a pristine Sc$_3$N@C$_{80}$ multilayer film.} \textbf{a} Excitons are created with a fs $3.2 \unit{eV}$ laser pulse and subsequently decay into energetically lower excitonic states. \textbf{b} The energy level diagram shows the energetic position of the excitons CT$_2$, CT$_1$ and S$_1$. \textbf{c} Population dynamics of the three excitonic levels shown in panel b. \textbf{d} Valence band structure of the studied Sc$_3$N@C$_{80}$ film for different pump-probe time delays (shown in green colors). The colored peaks show the contributions of the fitting model for the quasi-static ($\Delta t = -250 \unit{fs}$) spectrum (for other time steps, refer to Fig.~S3). \\
  \textbf{e} Transient change of the linewidth for the HOMO state (dots), fitted with a double exponential decay function convoluted with the pump-probe cross correlation curve (solid line).}
  \label{fgr:1}
\end{figure*}

Therefore, we focus on the tunability of the exciton dynamics of EMF materials by potassium (K) intercalation. As exemplary case, we have selected the tri-metallic nitride fullerene Sc$_3$N@C$_{80}$. It consists of a very robust C$_{80}$ carbon cage, which encapsulates three scandium (Sc) atoms coordinated to a central nitrogen (N) atom. This member of the EMF family is particularly interesting since the lowest unoccupied molecular orbital (LUMO) is not located at the carbon cage, but predominantly at the Sc$_3$N core \cite{Popov2008a}. This enables us not only to study the exciton dynamics at the carbon cage, but also to gain insights into potential cluster-cage charge transfer phenomena on ultrafast timescales.

%Here, we study the influence of the metal cluster in such an EMF on the exciton formation and decay dynamics triggered by light absorption. Especially molecules with the lowest unoccupied molecular orbital (LUMO) located at the metal core have the potential to show such an influence. A prominent example for EMFs with metal cluster located LUMO is the Sc$_3$N@C$_{80}$ molecule \cite{Popov2008a}.
%In the following, we present a fs time-resolved photoemission (trPES) study of a thin film of the prototypical endohedral metallofullerene Sc$_3$N@C$_{80}$ on a metal substrate. We show that the femtosecond electron dynamics following the optical excitation of the sample with a 390 nm UV light pulse are dominated by charge-transfer (CT) exciton formation and decay, transition to and trapping in Frenkel excitons. With K intercalation, we observe no trapping, suggesting that these dynamics can be strongly influenced by K intercalation.

In the following, we will show that K intercalation has a significant influence on the exciton dynamics of thin films of the prototypical endohedral metallofullerene Sc$_3$N@C$_{80}$. Using time-resolved photoemission with fs-XUV radiation, we find that the exciton dynamics of the pristine (undoped) molecular film after optical exciton with visible light are dominated by an exciton decay cascade starting from the thermalization of optically excited CT excitons to a trapping of the excited states in a long-lived Frenkel exciton. Upon K intercalation, the decay cascade ends after the thermalization of the CT excitons which dissolve on sub-$500 \unit{fs}$ timescales. We propose that this modification is caused by the dissociation of CT excitons due to the interaction with the K-atoms resulting in the generation of free charge carriers in homo-molecular materials on ultrafast timescales.

\section{Results and Discussion}
%\subsection{Femtosecond transient electron dynamics of the Sc$_3$N@C$_{80}$ multilayer}
\subsection{Ultrafast exciton dynamics of pristine Sc$_3$N@C$_{80}$ thin films}

We start our discussion with the ultrafast dynamics of the pristine Sc$_3$N@C$_{80}$ thin film ($\Theta_{\unit{Sc}_3\unit{N@C}_{80}} \approx 10 \unit{ML}$). The exciton dynamics were investigated in real-time using time-resolved photoelectron spectroscopy with fs-XUV radiation (experimental details can be found in the supplementary information). Crucially, the photon energy allows us to simultaneously monitor the excited state dynamics as well as the transient changes in the occupied valence levels of the entire molecular film. The latter can reveal characteristic transient spectral changes upon optical excitation, such as a transient linewidth broadening of all molecular valence states, which was recently identified as the transient spectroscopic signatures of CT excitons in molecular films \cite{Stadtmuller2019}.

%Upon excitation with the $3.2 \unit{eV}$ femtosecond laser pulse, we can distinguish two regimes, where different dynamics happen on the femto- to picosecond time scale. For energies above the highest occupied molecular orbital (HOMO) level, we can directly extract the \textit{population dynamics} from the spectral density shown in a 2D plot in figure \ref{fgr:1}a. Optical excitation of electrons from the HOMO into excited states leads to an immediate rise of spectral yield $2.8 \unit{eV}$ above the HOMO level (CT$_2$). This transient population is decaying over multiple steps and condenses in a state around $1.7 \unit{eV}$ above the HOMO level (S$_1$), with a lifetime of $\tau_{\unit{exc}} \gg 1 \unit{ps}$, as shown in figure \ref{fgr:1}a.

The exciton dynamics of the Sc$_3$N@C$_{80}$ film are shown as a two-dimensional plot of the time-dependent photoemission intensity in Fig.~\ref{fgr:1}a. After optical excitation with $3.2 \unit{eV}$ photons, we observe the instantaneous accumulation of spectral density $2.8 \unit{eV}$ above the highest occupied molecular orbital (HOMO) of the Sc$_3$N@C$_{80}$ molecule. This photoemission signal is identified as the CT$_2$ level, which is resonantly excited from the HOMO level (S$_0$ in Fig.~\ref{fgr:1}b) of the molecular film.
This transient population decays into the CT$_1$ level at $2.1 \unit{eV}$ above $E_{\unit{HOMO}}$ before becoming trapped in the excited state S$_1$ at $E-E_{\unit{HOMO}} = 1.7 \unit{eV}$ with a depopulation time of $\tau_{\unit{S}_1} \gg 1 \unit{ps}$. This decay cascade is summarized in the energy diagram in Fig.~\ref{fgr:1}b.

To quantify the exciton dynamics, we employed a fitting model with three Gaussian curves to model the time-dependent spectral density of the three states CT$_2$, CT$_1$ and S$_1$. During our fitting routine, we constrained the excited state energy and the linewidth (full width at half maximum, FWHM) of each excited state to a fixed value. The only free fitting parameter is the peak intensity (area) of each Gaussian curve, which reflects the population of each excitonic level (more details see supporting information).

The measured population dynamics of each excitonic state is shown in Fig.~\ref{fgr:1}c as colored dots. The solid lines of identical color represent the fitting model to determine the depopulation (decay) time of each excitonic state. The CT$_2$ exciton is formed instantaneously after the optical exciton (within our experimental uncertainty) and directly decays into the lower lying CT$_1$ exciton within $\tau_{\unit{CT}_2} = 45 \pm 5 \unit{fs}$ (black curve).
The decay dynamics of CT$_1$ can only be described by a double-exponential decay (see red curve in figure \ref{fgr:1}c) with two significantly different depopulation times of $\tau_{\unit{CT}_1,1} = 50 \pm 5 \unit{fs}$ and $\tau_{\unit{CT}_1,2} = 2.5 \pm 1.0 \unit{ps}$. Subsequently, the population of the S$_1$ exciton increases within the first $200 \unit{fs}$ before staying almost constant for all times investigated in our experiment.
This points to an excitonic lifetime of the S$_1$ level of $\tau_{\unit{S}_1} \gg 10 \unit{ps}$.

The double-exponential decay trace of CT$_1$ suggests the existence of at least two excitonic states in this excited state range which could not be separated in our data. In fact, the spectral shape of the excited states changes continuously during the transition from the CT$_2$ to the CT$_1$ exciton. However, no reliable fitting results were obtained when allowing a continuous change of the spectral line shape of the CT$_2$ exciton. Therefore, we had to limit our fitting model to three excitonic states CT$_2$, CT$_1$ and S$_1$ at fixed excited state energies.

The spatial charge distribution of the three excitonic states can be revealed by analyzing the occupied valence states' dynamics. Figure \ref{fgr:1}d shows the general trend of the optically induced band structure dynamics of the occupied molecular levels by displaying selected EDCs for characteristic time steps of the previously observed exciton dynamics in the excited states. First, before optical excitation ($\Delta t = -250 \unit{fs}$), the quasi-static spectral line-shape of the HOMO level as well as the lower lying molecular orbitals (HOMO-X) reflect perfectly the spectroscopic findings of previous studies of Sc$_3$N@C$_{80}$ multilayer films on different surfaces \cite{Seidel2018,Alvarez2002}.
Upon optical excitation and creation of excitons, the linewidth of all molecular features increases before transforming back to its original state. The relaxation dynamics are significantly slower than the excitation and occur only within several hundreds of fs. This transient broadening of all molecular valence orbitals reflects the many-body response of the molecular material to the optically induced excitons, caused by electronic correlations and dielectric screening. Even more importantly, the transient broadening was recently identified as the spectroscopic signature of CT excitons in the molecular material \cite{Stadtmuller2019} and hence clearly hints an at least partial CT character of the optically excited excitons in the Sc$_3$N@C$_{80}$ film.

The transient changes of the spectral linewidth were quantified by applying a dedicated fitting routine (see SI for a detailed description of the valence orbital fitting routine). The extracted transient evolution of the linewidth of the HOMO level is shown in Fig.~\ref{fgr:1}e. We find an instant rise of the transient linewidth after optical excitation within our fs temporal resolution \cite{Eich2014} followed by a double-exponential decay with two significantly different time constants.
These time constants are $\tau_{\unit{TB}_2} = 90 \pm 15 \unit{fs}$ and $\tau_{\unit{TB}_1} = 2.5 \pm 0.5 \unit{ps}$.% with a maximum amplitude of the transient broadening of $\Delta \unit{FWHM}_{\unit{max,TB}_2} = 45 \pm 2 \unit{meV}$ and $\Delta \unit{FWHM} _{\unit{max,TB}_1} = 11 \pm 2 \unit{meV}$.
These time constants match almost perfectly the depopulation times of the CT$_2$ and the CT$_1$ excitons, but not the one of the S$_1$ exciton. Therefore, only the CT$_2$ and CT$_1$ excitons reveal partial CT character.

Together, our findings for exciton population dynamics (excited states) and the transient charge character of the excitons (valence states) provides a clear picture of the exciton dynamics in the pristine Sc$_3$N@C$_{80}$ film. Optical exciton with $3.2 \unit{eV}$ photons results in the instantaneous formation of hot excitons with at least partial CT character (referred to as CT$_2$). These CT$_2$ excitons thermalize within $100 \unit{fs}$ and transform continuously into the CT$_1$ excitons with lower excited state energy.
In a second step, the thermalized CT$_1$ excitons decay further into the S$_1$ exciton with dominant Frenkel exciton-like character where they become trapped for at least $10 \unit{ps}$. Interestingly, the energy of the of S$_1$ exciton ($E-E_{\unit{HOMO}} = 1.7 \unit{eV}$) is still significantly larger than the energy of the LUMO-derived molecular state inside the carbon cage \cite{Seidel2018,Alvarez2002}.
This suggests that the exciton dynamics of all three excitonic states CT$_2$, CT$_1$ and S$_1$ only involve excited states that are located at the carbon cage, but not at the Sc$_3$N core. It is hence very similar to the exciton dynamics of the prototypical fullerene C$_{60}$ \cite{Stadtmuller2019}.

\begin{figure*}[ht]
  \includegraphics[width=110mm]{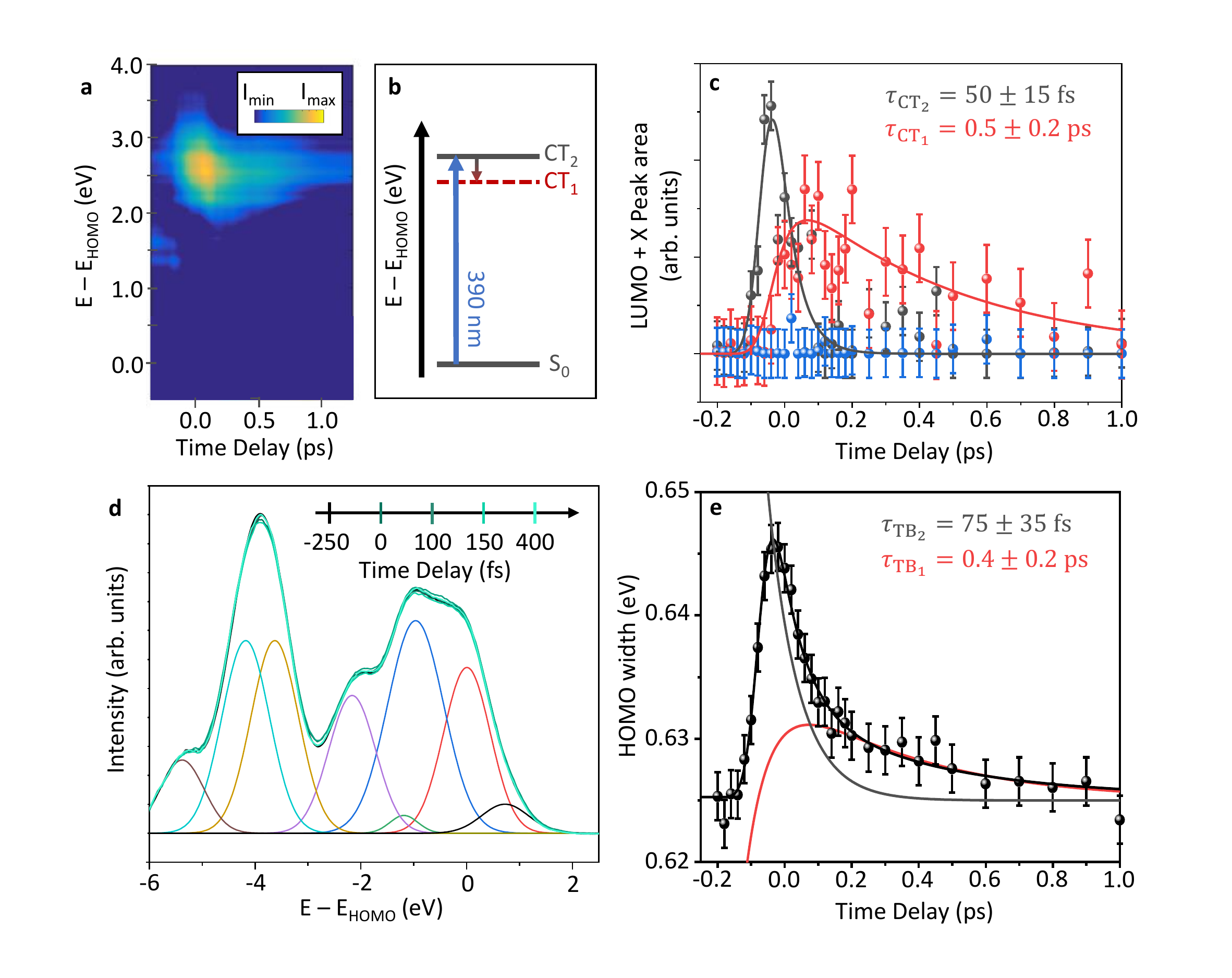}
  \caption{\textit{Electron dynamics in a K-intercalated K$_1$(Sc$_3$N@C$_{80}$) multilayer film.} \textbf{a} Excitons are created with a fs $3.2 \unit{eV}$ laser pulse and subsequently decay into energetically lower excitonic states. \textbf{b} The energy level diagram shows the energetic position of the excitons CT$_2$ and CT$_1$. \textbf{c} Population dynamics of the excitonic levels shown in panel b.
  Besides CT$_2$ and CT$_1$, the excitonic state S$_1$, populated during the decay cascade of the pristine Sc$_3$N@C$_{80}$ film (see fig.~\ref{fgr:1}), has been included in the fitting routine (blue dots), but is not populated for the K-intercalated case (compare Fig. S2).
  \textbf{d} Valence band structure of the studied K-intercalated Sc$_3$N@C$_{80}$ film for different pump-probe time delays (shown in green colors). The colored peaks show the contributions of the fitting model for the quasi-static spectrum at $\Delta t = -250 \unit{fs}$ (for other time steps, refer to Fig. S4). \textbf{e} Transient change of the linewidth for the HOMO state (dots), fitted with a double exponential decay function convoluted with the pump-probe cross correlation curve (solid line).}
  \label{fgr:2}
\end{figure*}

\subsection{Exciton dynamics of a K-intercalated K$_1$(Sc$_3$N@C$_{80}$) thin film}
We now turn to the modification of the exciton dynamics of the Sc$_3$N@C$_{80}$ film after K intercalation and start with a K concentration with one K-atom per Sc$_3$N@C$_{80}$ molecule ($x = 1$).

First of all, K intercalation of the pristine Sc$_3$N@C$_{80}$ thin film results in a spectral broadening of the valence band structure coinciding with a work function decrease of $\Delta \Phi = 0.5 \pm 0.05 \unit{eV}$ \cite{Alvarez2002}. This spectral broadening has its origin in the lifting of the degeneracy of the $\pi$-orbitals located at the C$_{80}$ cage, which is caused by the reduced cage symmetry of the doped fullerene cage \cite{Harigaya1992}.
The most important change, however, is the population of the LUMO-derived state located at the Sc$_3$N core.
It becomes occupied by the CT from the K-dopant atom to the Sc$_3$N@C$_{80}$ and leads to a new spectroscopic feature in the molecular valence band structure located $0.85 \unit{eV}$ above $E_{\unit{HOMO}}$ \cite{Popov2008a, Alvarez2002}. It is marked by a black Gaussian curve in the spectra in Fig.~\ref{fgr:2}d.
%This change of population inside the carbon cage causes a decrease of the \textit{d}$_{\unit{Sc}} - \pi_{\unit{cage}}$ wave function overlap, with the result that the overall Sc3\textit{d} population is not significantly altered.

Besides these spectral changes, K intercalation has a strong effect on the optically induced exciton dynamics of the K$_1$(Sc$_3$N@C$_{80}$) thin film which is summarized in Fig.~\ref{fgr:2}. The measured excited state dynamics are shown in Fig.~\ref{fgr:2}a, the corresponding energy level diagram in Fig.~\ref{fgr:2}b. Optical excitation results in the  instantaneous formation of hot excitons $2.8 \unit{eV}$ above $E_{\unit{HOMO}}$ (CT$_2$). Subsequently, these hot CT$_2$ excitons thermalize and continuously decay into the excitonic level CT$_1$ at lower energies.
The exciton population is trapped in the CT$_1$ exciton for a hew hundreds of fs before vanishing from the excited states. In particular, we do not observe any indication for a long-lived trapping of excitons in the S$_1$ level as discussed beforehand for the pristine Sc$_3$N@C$_{80}$ films.

The time scales of the exciton decay cascade can be determined from the transient population of the CT$_2$ and the CT$_1$ state shown in Fig.~\ref{fgr:2}c. These values were extracted from the time-resolved photoemission data by using Gaussian curves at fixed excited state energies as discussed above. The best fitting result was obtained for a slightly larger excited state energy of Gaussian peak modelling the thermalized CT$_1$ excitons. Interestingly, no population could be detected for the energy position of the S$_1$ exciton (blue data points).
Using exponential decay functions, we observe depopulation times of $\tau_{\unit{CT}_2,\unit{K}} = 50 \pm 15 \unit{fs}$  and $\tau_{\unit{CT}_1,\unit{K}} = 0.5 \pm 0.2 \unit{ps}$ for the CT$_2$ and CT$_1$ excitons.
The depopulation time for the CT$_1$ state is hence almost one order of magnitude smaller after K intercalation than in the pristine Sc$_3$N@C$_{80}$ film. These time scales can be directly compared to each other as the K intercalation study was conducted on the same sample as the study of the exciton dynamics of the pristine  Sc$_3$N@C$_{80}$ film. This suggests a new and significantly faster decay (or energy dissipation) mechanism for the thermalized CT$_1$ excitons due to the presence of K impurities.
\begin{figure*}[th]
  \includegraphics[width=130mm]{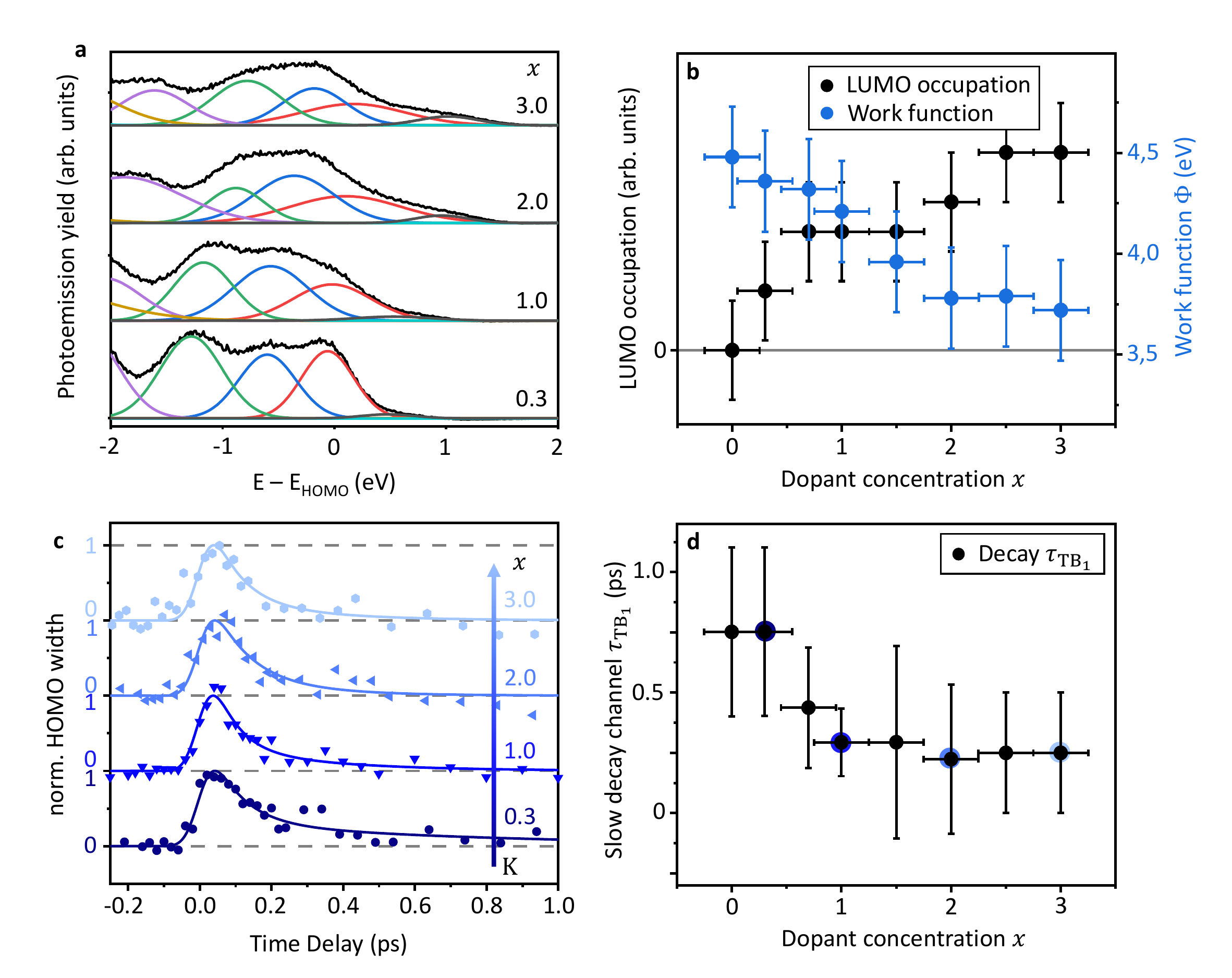}
  \caption{\textit{K concentration-dependent intercalation study of a K$_x$(Sc$_3$N@C$_{80}$) multilayer film.} \textbf{a} Valence band photoemission spectra of K$_x$(Sc$_3$N@C$_{80}$) for selected intercalation steps (see supplementary figure S5b for all datasets). The K concentration $x$ has been estimated by comparison of  valence band photoemission spectra with literature \cite{Alvarez2002} (compare figure S5a). The line-shape of the Gaussian curves show the contributions to the valence band photoemission spectra of K$_x$(Sc$_3$N@C$_{80}$). \\
  \textbf{b} Increasing LUMO occupation (black dots; left vertical scale) and decreasing work function (blue dots; right vertical scale) extracted from the fit of the valence orbital photoemission spectra in \textbf{a} as a function of dopant concentration $x$.
  \textbf{c} Normalized transient HOMO linewidth for selected intercalation steps (shifted vertically for better comparability). All data sets are shown in supplementary figure S5c. \textbf{d} Extracted decay time $\tau_{\unit{TB}_1}$ for different dopant concentrations $x$. The decay time decreases for K concentrations $x > 0.3$. Colored data points are extracted from the respective graph in \textbf{c}. All data sets used for extraction of the shown data are shown in supplementary figure S5c.}
  \label{fgr:3}
\end{figure*}

These changes are also reflected in the transient linewidth broadening $\Delta \unit{FWHM}$ of the valence states. EDCs at selected time delays are shown in Fig.~\ref{fgr:2}d, the extracted transient linewidth broadening of the HOMO is illustrated in Fig.~\ref{fgr:2}e. Similar to the pristine Sc$_3$N@C$_{80}$ film, there is an immediate broadening of the HOMO upon the excitation followed by a double-exponential decay back to the initial spectral linewidth.
The decay times extracted from the double-exponential decay fit are $\tau_{\unit{TB}_2,\unit{K}} = 75 \pm 35 \unit{fs}$ and $\tau_{\unit{TB}_1,\unit{K}} = 0.4 \pm 0.2 \unit{ps}$, respectively. Both decay times match the depopulation times of the CT$_2$ and CT$_1$ excitons in the excited states, demonstrating that both types of excitons still reveal dominant CT character, even after K intercalation.

Hence, K intercalation results in a significantly faster decay of the thermalized charge-transfer excitons CT$_1$ compared to the CT$_1$ excitons of the pristine Sc$_3$N@C$_{80}$ thin films. Crucially, these CT$_1$ excitons do not decay into lower lying excitonic levels, such as the S$_1$ exciton for the pristine film, but disappear completely from the excited state spectra. These observations could either be attributed to a direct recombination of the electron and hole of the CT$_1$ exciton, i.e., to a direct decay of the CT$_1$ exciton into the ground state, or to an efficient exciton dissociation and the generation of free carriers. In the following, we tackle this question by studying the correlation between the exciton dynamics of K-intercalated Sc$_3$N@C$_{80}$ thin films for various K concentrations.

\subsection{K concentration-dependent exciton dynamics of K$_x$(Sc$_3$N@C$_{80}$) thin films}
The K concentration-dependent exciton dynamics of the Sc$_3$N@C$_{80}$ thin film are shown in Fig.~\ref{fgr:3}. The K concentration of each intercalation step was identified by the characteristic line-shape of the (static) valence band structure \cite{Alvarez2002} which is shown in Fig.~\ref{fgr:3}a for selected K concentrations $x$.
The continuous broadening of the valence band structure with increasing K concentration can be attributed to a K-induced structural distortion of the C$_{80}$ cage coinciding with a change in the cage symmetry and a lifting of the energy degeneracy of the molecular valence orbitals \cite{Harigaya1992,Popov2008a,Alvarez2002}. In addition, increasing the K concentration results in a gradual population of the LUMO-derived state at the Sc$_3$N core due to a charge transfer from the K atoms into the molecule.
The amount of charge transfer can be estimated by the spectral weight of the corresponding LUMO-derived feature in the valence band spectra which emerges at $0.85 \unit{eV}$ above $E_{\unit{HOMO}}$ (black Gaussian curve in Fig.~\ref{fgr:2}d and Fig.~\ref{fgr:3}a). It is shown in Fig.~\ref{fgr:3}b as black data points together with the K concentration-dependent decrease of the work function.

For each intercalation step $x$, the excited states and the transient evolution of the valence states were studied by time-resolved photoemission. Here, we focus our discussion on the K-induced modifications of the population dynamics of the excitons with dominant CT character, which dominate the first steps of the exciton thermalization and decay process in fullerene materials. Their dynamics are fully reflected in the transient linewidth broadening of the molecular valence bands which is shown for the exemplary case of the HOMO level in Fig.~\ref{fgr:3}c for selected K concentrations $x$. The transient linewidth of the HOMO has been normalized to its linewidth before optical excitation for better comparability of the K-dependent transient linewidth traces. All transient linewidth traces were analyzed by the double-exponential decay model presented earlier and resulted in K concentration-dependent decay times $\tau_{\unit{TB}_2}(x)$ and $\tau_{\unit{TB}_1}(x)$. Within our experimental uncertainty, the fast decay time $\tau_{\unit{TB}_2}(x)$, connected to the initial optical excitation, stays constant for all K concentrations, which suggests that the thermalization dynamics of the optically excited, hot CT$_2$ excitons do not dependent on the amount and density of the K atoms within the Sc$_3$N@C$_{80}$ films.

In contrast to the dynamics of the hot CT$_2$ excitons, we find two characteristic intercalation regimes for the K-dependent dynamics of the more thermalized CT$_1$ excitons which are shown in Fig.~\ref{fgr:3}d. In the first regime between $0.3<x<1$, the depopulation times $\tau_{\unit{TB}_1}(x)=\tau_{\unit{CT}_1}(x)$ of CT$_1$ change quantitatively and decrease from $0.75 \pm 0.25 \unit{ps}$ to $0.25 \pm 0.2 \unit{ps}$, i.e., the average speed of the exciton decay process increases with increasing K concentration.
In parallel, we observe a reduction of the spectral yield of the S$_1$ exciton, which completely vanishes for K concentrations $x \geq 1$.
In this regime, we observe a saturation of $\tau_{\unit{TB}_1}(x \geq 1)$ at $0.25 \pm 0.2 \unit{ps}$ and the disappearance of any signature of the S$_1$ excitonic level in the excited states. Hence, there is a characteristic and qualitative modification of the exciton dynamics of the K$_x$(Sc$_3$N@C$_{80}$) films above the critical K concentration of $x=1$. Therefore, the exciton dynamics of the K$_1$(Sc$_3$N@C$_{80}$) film discussed above are representative for all K concentrations in this regime ($x \geq 1$).

Interestingly, the transition of the exciton dynamics occur for a K concentration with one K atom per Sc$_3$N@C$_{80}$ and not at $x=2$, where the lowest LUMO state, located at the core, would be completely occupied (see Fig.~\ref{fgr:3}d). This suggests that the population of the LUMO level is not the decisive factor in the qualitative transition of the exciton dynamics since population-induced blocking of final states for the exciton relaxation process can only occur efficiently for fully occupied molecular orbitals. Instead, we propose that the direct decay or dissociation of the CT$_1$ is most likely triggered by the spatial vicinity of the CT exciton and the K atom.

\section{Conclusion}
Our experimental findings have demonstrated the decisive influence of K intercalation on the exciton decay processes of endohedral metallofullerenes. The exciton dynamics of the pristine Sc$_3$N@C$_{80}$ thin film reveal the typical exciton dynamics of fullerene thin films \cite{Jacquemin1998,Rosenfeldt2010,Causa2018,Stadtmuller2019}, i.e., the wave functions of the excitonic level are mainly localized on the carbon cage.
First, optical excitation with visible light results in the formation of hot excitons with CT character (CT$_2$). These hot CT$_2$ excitons relax within sub-$100 \unit{fs}$ into thermalized CT excitons (CT$_1$) with significantly longer depopulation times in the order of $1 \unit{ps}$.
Finally, the excitons relax into a Frenkel-like excitonic level S$_1$ in which they become trapped for several ps.

After K intercalation, the exciton dynamics of the molecular material are strongly modified above a characteristic step, corresponding to a distinct K concentration regime. For moderate doping concentrations with less than one K atom per Sc$_3$N@C$_{80}$, we observe a continuous reduction of the depopulation times of the CT$_1$ and a decreasing maximal population of the S$_1$ exciton which increasing K concentration. For larger K concentrations with at least one K atom per Sc$_3$N@C$_{80}$ molecule, the CT$_1$ excitons no longer decay into the energetically lower S$_1$ state, but directly dissolve or dissociate with a constant depopulation time below $0.5 \unit{ps}$.

Most importantly, the characteristic transition between these two exciton decay processes occurs below one K atom per Sc$_3$N@C$_{80}$ molecule.
This suggests that either the direct spatial vicinity between the CT excitons and the K dopants or a minimum charge transfer between the K dopants and the Sc$_3$N@C$_{80}$ molecules is required for the modification of the excitation dynamics.

Charge transfer into the Sc$_3$N@C$_{80}$ molecule results in an occupation of the 3\textit{d} orbitals of the Sc atoms in the metal core coinciding with a reducing of the wave function overlap between the Sc$_3$N cluster and the C$_{80}$ cage. This can lead to a change of the cage symmetry of the pristine Sc$_3$N@C$_{80}$ \cite{Popov2008a,Alvarez2002} and to lattice distortions of the molecular films \cite{Harigaya1992} upon K intercalation. This could influence the excited state decay channels for carbide-based clusterfullerenes \cite{Wu2015}.
However, these modifications of the structural and electronic properties of the molecules still evolve, even for K concentrations larger than one K atom per Sc$_3$N@C$_{80}$ molecule, as is clearly visible in our photoemission data in Fig.~\ref{fgr:3}. This is shown, for instance, by the continuous modifications of the molecular valence band structure for K concentrations $x \geq 1$.

Therefore, our findings indicate that the interaction between CT excitons and K-dopant atoms are responsible for the qualitative transition in the exciton decay dynamics. The charge-transfer process between the K atom and the Sc$_3$N@C$_{80}$ molecule leads to an (at least partial) ionic character of the K-dopants in the molecular film. As a result, the dynamics of the CT excitons in the vicinity of a K ion are severely altered by the interaction of the microscopic charge distribution of the CT excitons and the Coulomb potential of the K ions. This additional interaction opens a new relaxation pathway either towards the recombination of the electron and hole of the exciton or towards exciton dissociation and the generation of free charge carriers. An enhanced exciton decay via recombination is more likely in molecular materials with enhanced structural or charge order which, for instance, occurs by reducing the thermal motion of the molecules using e.g. temperature \cite{Dresselhaus1996,Janner1995}. This is clearly different in our case, where K intercalation results in the lattice distortions of the Sc$_3$N@C$_{80}$ molecule and the increase of K defect sites that break the translation symmetry of the molecular layer. Therefore, we propose that K interaction significantly enhances the speed and efficiency of the free charge carrier generation via exciton dissociation. Possibly, the ionic K dopants are able to capture the electron of the excitation via Coulomb attraction resulting in the generation of free holes in the Sc$_3$N@C$_{80}$ film.

Finally, we would like to point out that our model can also explain the independence of the depopulation time of the CT$_1$ exciton on K intercalation for low intercalation concentrations. For low K concentrations, the probability to create CT excitons in direct vicinity of a K atom is rather low due to the random distribution of the K dopants in the molecular film and the stochastical process of the optical excitation. Hence, only a small number of CT excitons can dissociate via the ultrafast K-mediated relaxation channel while the other CT excitons decay with the intrinsic depopulation time of the pristine films before becoming trapped in the S$_1$ state. Our photoemission experiment spatially averages over all these processes and the recorded signal is composed of both decay processes according to their relative contributions. With increasing K concentration, the relative contribution of the direct exciton dissociation process increases and the average depopulation time constant of CT$_1$ decreases until is saturates at $x=1$. At this point, all CT excitons decay via the much faster relaxation channel mediated by the K intercalation-induced charge defects.

In conclusion, our investigation of the exciton dynamics in K-intercalated endohedral metallofullerenes has demonstrated the decisive role of alkali metal intercalation on the ultrafast exciton dynamics of molecular materials. We uncovered a K-induced ultrafast relaxation mechanism that prevents the trapping of excitons in long-lived excitonic states and results, most likely, in the generation of free charge carriers on ultrafast, sub-$500 \unit{fs}$ timescales. We therefore propose alkali metal doping of molecular films as a novel approach to enhance the light-to-charge carrier conversion efficiency in photovoltaic materials that could potentially pave the way towards the next generation of molecular-based light-harvesting applications with superior performance.

\begin{acknowledgments}
The research leading to these results was financially supported by the Deutsche Forschungsgemeinschaft (DFG, SFB/TRR 88 “Cooperative Effects in Homo- and Heterometallic Complexes (3MET)” Project C9). Furthermore, B.S. and S.E. acknowledge financial support from the Graduate School of Excellence Mainz (Excellence initiative DFG/GSC 266). This work is supported by the European Research Council (Grant 725767-hyControl).
\end{acknowledgments}

%\newpage

%\onecolumn
%\bibliography{Bibliography/C80_lib}% Produces the bibliography via BibTeX.
%\includepdf[pages={{},1,{},2,{},3,{},4,{},5,{},6,{},7,{},8,{},9,{}}]{Sc3NC80_SI.pdf}
\end{document}

%\title{Control of ultrafast exciton decay channels via Potassium intercalation into the endohedral metallofullerene semiconductor Sc$_3$N@C$_{80}$}

\title{Ultrafast charge carrier separation in Potassium-intercalated endohedral metallofullerene Sc$_3$N@C$_{80}$ thin films (Supplementary Information)}

\author{Sebastian Emmerich}
%\email{emmerich@physik.uni-kl.de}
\affiliation{University of Kaiserslautern and Research Center OPTIMAS, Erwin-Schr\"odinger-Stra\ss{}e 46, 67663 Kaiserslautern, Germany}
\author{Sebastian Hedwig}
\affiliation{University of Kaiserslautern and Research Center OPTIMAS, Erwin-Schr\"odinger-Stra\ss{}e 46, 67663 Kaiserslautern, Germany}
\author{Mirko Cinchetti}
\affiliation{Experimentelle Physik VI, Technische Universit\"at Dortmund, 44221 Dortmund, Germany}
\author{Benjamin Stadtm\"uller}
\affiliation{University of Kaiserslautern and Research Center OPTIMAS, Erwin-Schr\"odinger-Stra\ss{}e 46, 67663 Kaiserslautern, Germany}
\author{Martin Aeschlimann}
\affiliation{University of Kaiserslautern and Research Center OPTIMAS, Erwin-Schr\"odinger-Stra\ss{}e 46, 67663 Kaiserslautern, Germany}
%\date{February 2020}%

\maketitle
\tableofcontents

\newpage
\newpage
\section{Supplementary Figures}
\begin{figure*}[ht]
  \includegraphics[width=180mm]{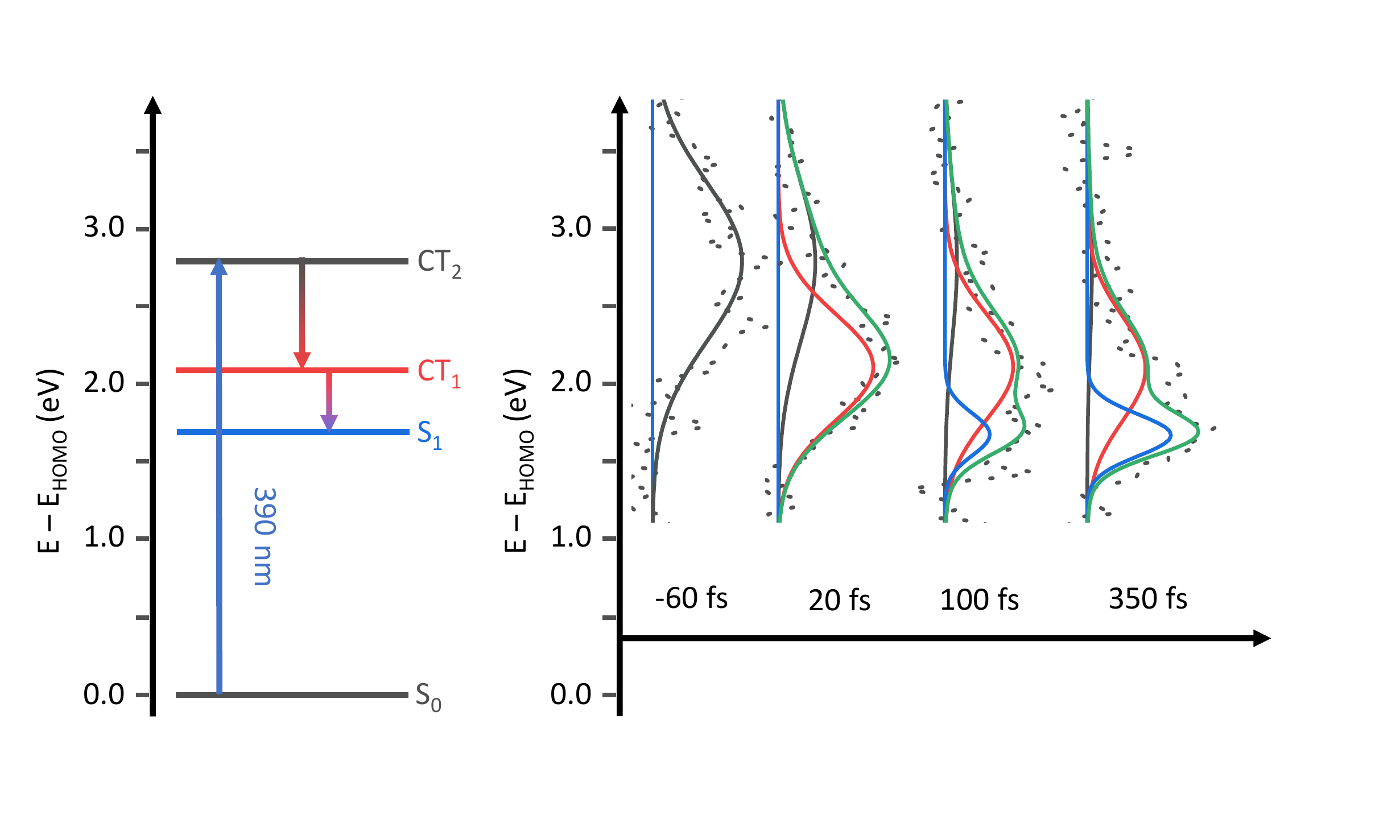}
  \caption{\textit{Excited state dynamics of the pristine Sc$_3$N@C$_{80}$ film for selected time delay steps, complementing figure 1a-c of the main text.} On the \textit{left} side, the energy level diagram for the pristine Sc$_3$N@C$_{80}$ film is shown, as presented and discussed in the main text. On the \textit{right} side, energy cuts at selected pump-probe time delays are plotted on the same energy axis as the energy level diagram on the left side, in arbitrary units (\textit{black dots}). The overall fitting function and the individual Gaussian peaks used for the fit are included as colored lines. While position and linewidth were constrained to constant values, the peak area, representing the transient population of each excited state, is the only free parameter of the fitting routine. The resulting transient population (area) of all three Gaussian curves are plotted in figure 1c of the main text.
  The Gaussian curve modelling CT$_2$ (\textit{solid black}) has a width of $\Delta E = 0.75 \unit{eV}$ and a position of $E-E_{\unit{HOMO}} = 2.8 \unit{eV}$, the second Gaussian, modelling CT$_1$ (\textit{solid red}) has a width of $\Delta E = 0.5 \unit{eV}$ and a position of $E-E_{\unit{HOMO}} = 2.1 \unit{eV}$ and the Gaussian modelling S$_1$ (\textit{solid blue}) has a width of $\Delta E = 0.2 \unit{eV}$ and a position of $E-E_{\unit{HOMO}} = 1.7 \unit{eV}$.}
  \label{fgr:S1}
\end{figure*}

\newpage

\begin{figure*}[ht]
  \includegraphics[width=180mm]{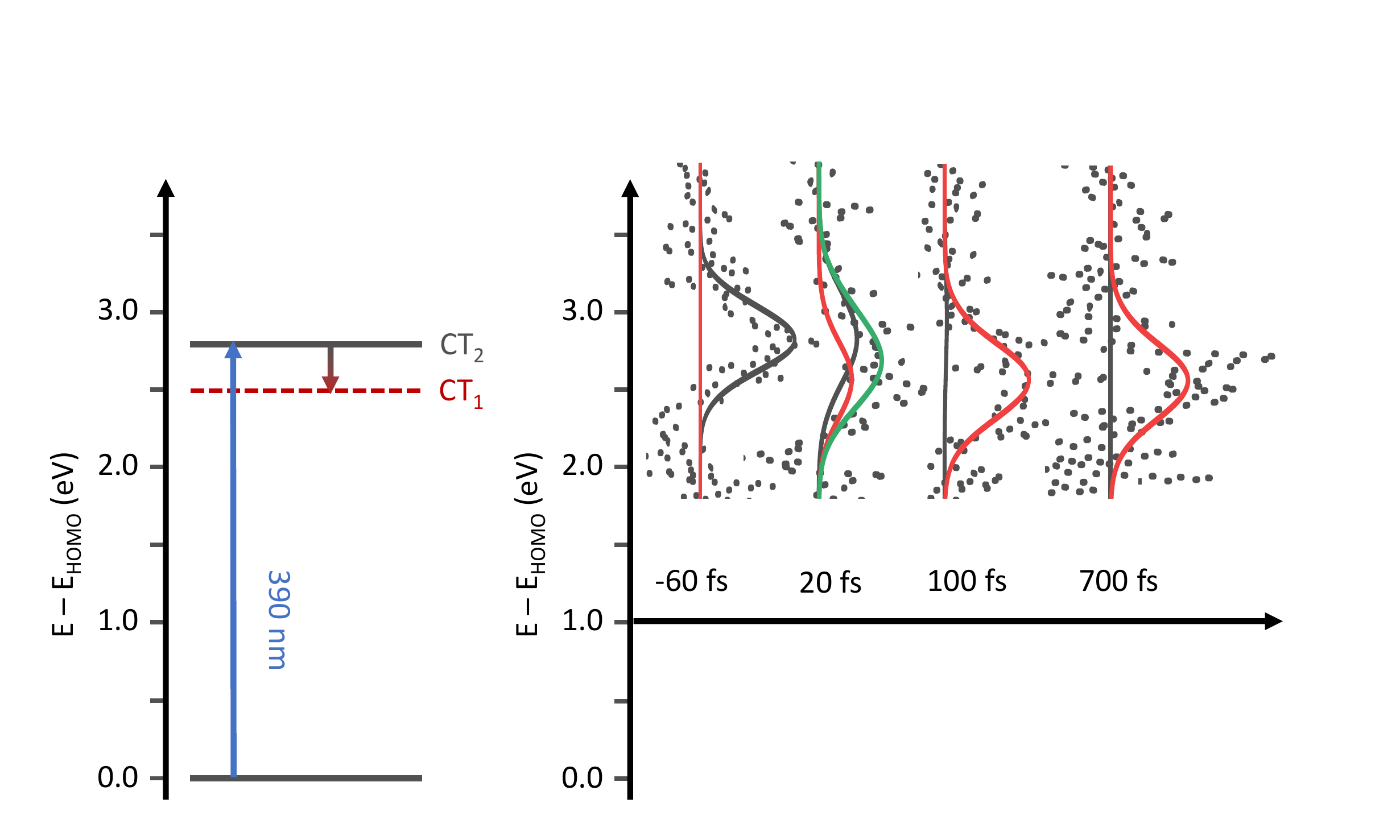}
  \caption{\textit{Excited state dynamics of the K intercalated K$_1$(Sc$_3$N@C$_{80}$) film for distinct time delay steps, complementing figure 2a-c of the main text.} On the \textit{left} side, the energy level diagram for the K intercalated K$_1$(Sc$_3$N@C$_{80}$) film is shown, as presented and discussed in the main text. On the \textit{right} side, energy cuts at selected pump-probe time delays are plotted on the same energy axis as the energy level diagram on the left side, in arbitrary units (\textit{black dots}). The overall fitting function and the individual Gaussian peaks used for the fit are included as colored lines. While position and linewidth were constrained to constant values, the peak area, representing the transient population of each excited state, is the only free parameter of the fitting routine. The resulting transient population (area) of all three Gaussian curves are plotted in figure 2c of the main text.
  The Gaussian modelling CT$_2$ (\textit{solid black}) has a width of $\Delta E = 0.7 \unit{eV}$ and a position of $E-E_{\unit{HOMO}} = 2.8 \unit{eV}$, the second Gaussian, modelling CT$_1$ (\textit{solid red}) has a width of $\Delta E = 0.6 \unit{eV}$ and a position of $E-E_{\unit{HOMO}} = 2.45 \unit{eV}$. }
  \label{fgr:S2}
\end{figure*}

\newpage

\begin{figure*}[ht]
  \includegraphics[width=170mm]{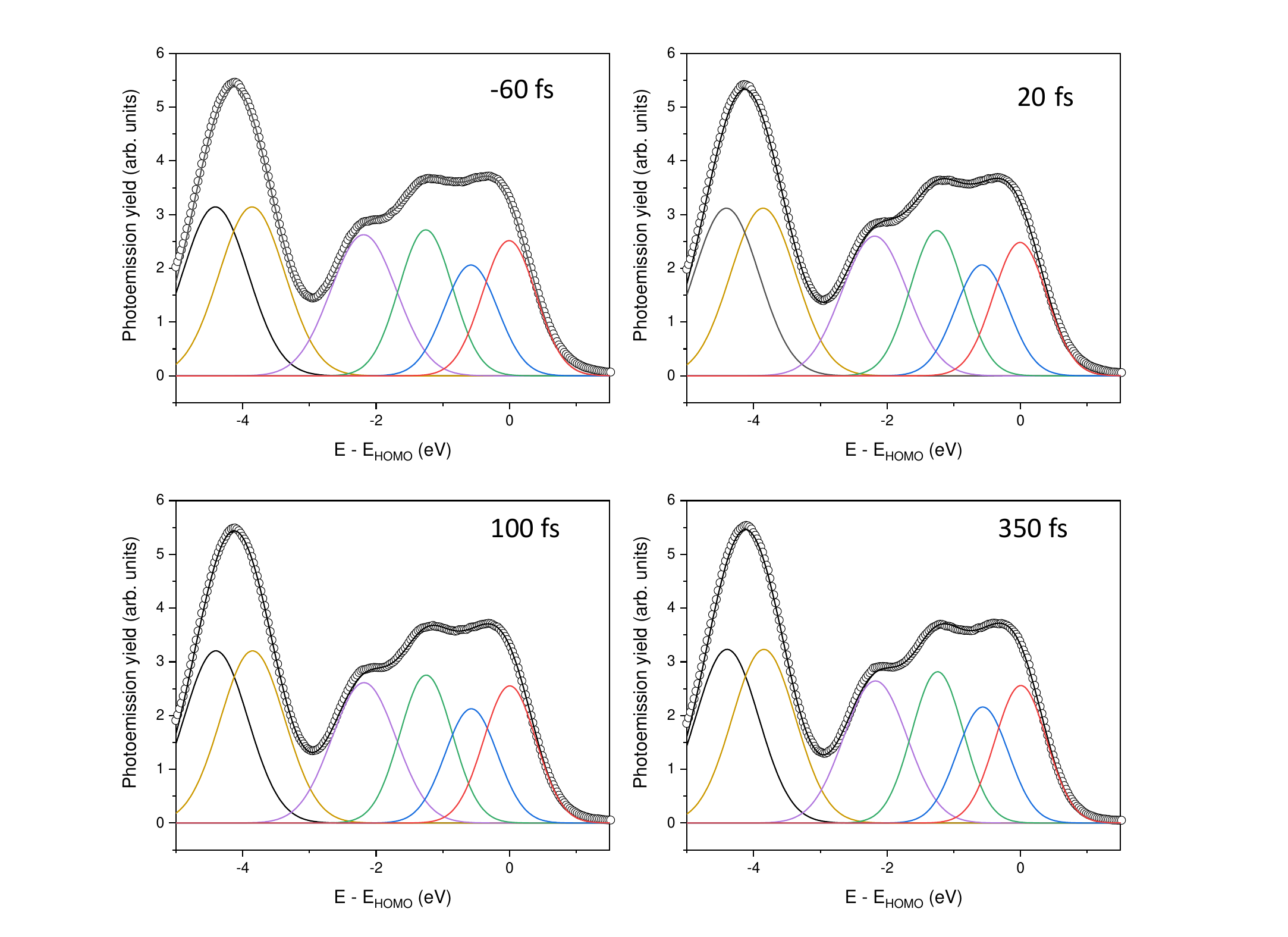}
  \caption{\textit{Fitting model for the transient linewidth broadening of the valence band structure at selected time delays, complementing figure 1d of the main text.} The spectra have been recorded using the snapshot mode of our detector ($E_{\unit{pass}} = 100 \unit{eV}$) together with our femtosecond-XUV light source ($h \nu = 22.2 \unit{eV}$). Each datapoint is calculated from a 2D image with axis of kinetic electron energy $E_{\unit{kin}}$ versus electron emission angle $\Theta$ for a fixed pump-probe delay. Changing the pump-probe time delay $\Delta t$, we acquire our 3D-data stack $I(E_{\unit{kin}},\Theta,\Delta t)$. To obtain the shown 1D plots, we extract a cut along the energy and average the information in a small angle region of around $3^{\circ}$. The fitting routine is a two-step process, starting with an exponential background subtraction, which has already been performed for the data shown above. The second step consists of analyzing the spectral line-shape of the valence band, using the least number of Gaussian peaks needed for a decent description of the spectrum.
  After optimizing the absolute position, width and area of each Gaussian peak to match the spectral line-shape of the quasi-static spectrum at $\Delta t = -250 \unit{fs}$, the relative position, area and width of all Gaussian peaks are constrained to these values during the fitting process. The result of the fit of the time-dependent data are only three free parameters, namely one free area, position and width. The underlying assumption of this approximation is that all valence states will react similarly to the perturbation induced by absorption of UV light, as demonstrated recently in our previous work \cite{Stadtmuller2019}. While the time evolution of the area and position stay constant in time within error bars, the spectral linewidth changes over time, as shown in figure 1e of the main manuscript.  }
  \label{fgr:S3}
\end{figure*}

\newpage

\begin{figure*}[ht]
  \includegraphics[width=170mm]{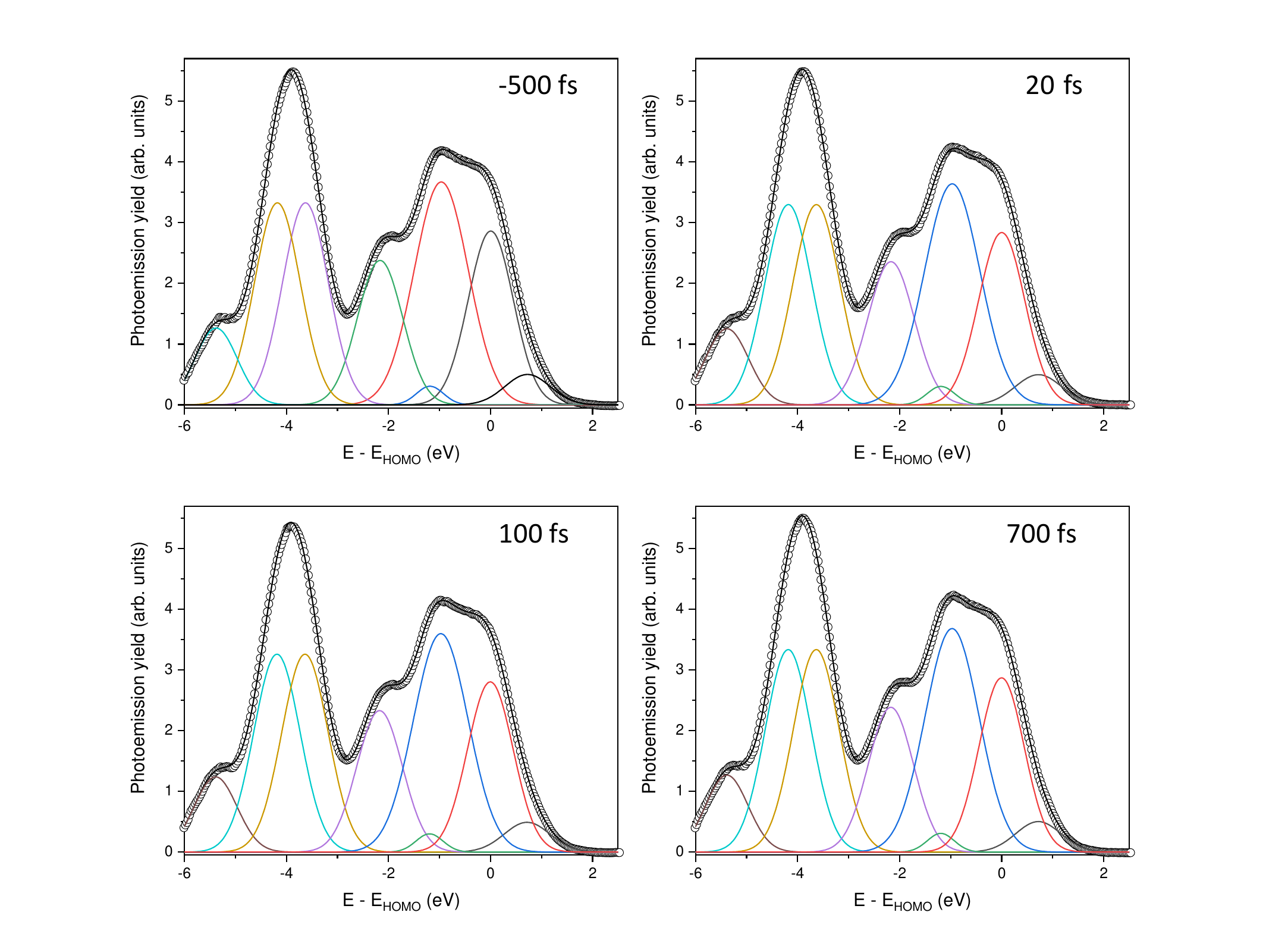}
  \caption{\textit{Valence orbital fit of the K intercalated K$_1$(Sc$_3$N@C$_{80}$) film for distinct time delay steps, complementing figure 2d of the main text.} For details about the fitting procedure for the valence band photoemission spectra shown here, see caption of figure \ref{fgr:S3} of the supplementary information.}
  \label{fgr:S4}
\end{figure*}

\newpage

\begin{figure*}[ht]
  \includegraphics[width=180mm]{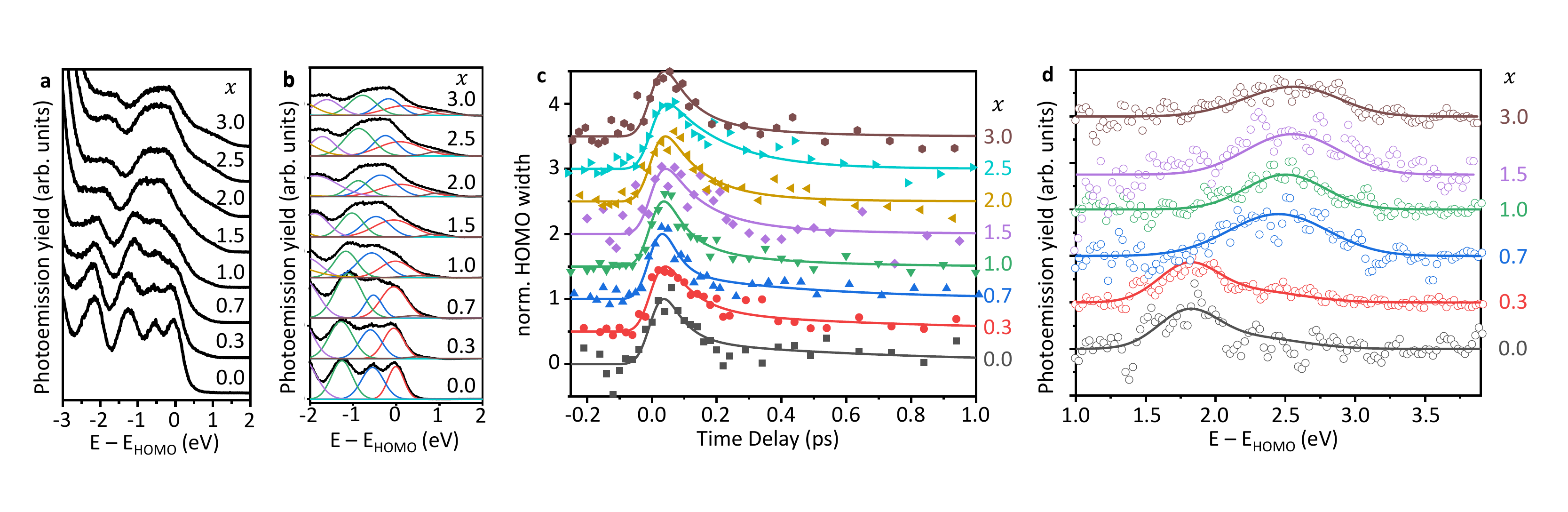}
  \caption{\textit{Static valence band spectra, transient linewidth broadening and excited state population for different intercalation steps $x$ of the K$_x$(Sc$_3$N@C$_{80}$) film at a time delay of $300 \unit{fs}$, complementing figure 3 of the main text.}
  \textbf{a} Static valence band spectra taken for each intercalation step $x$, plotted in a similar way as previously by Alvarez \textit{et al.}~\cite{Alvarez2002}. The spectral shape of these valence band photoemission data sets was used to calibrate the K concentration $x$, the absolute intercalation density per Sc$_3$N@C$_{80}$ molecule.
  \textbf{b} UPS valence spectra with Gaussian curves used to model the valence states, as plotted in figure 3a of the main text, but for each intercalation step.
  \textbf{c} Normalized transient linewidth broadening of HOMO, extracted from the fs pump-probe spectroscopy valence spectra for all K concentrations $x$, as plotted in figure 3c of the main text.
  \textbf{d} Excited state spectra recorded at $\Delta t = 300 \unit{fs}$ for the pristine ($x = 0$) and K-intercalated film with different K concentrations $x$. The spectra are vertically shifted for better visibility. At $\Delta t = 300 \unit{fs}$, the excited state spectra of the pristine and marginally K-intercalated film already reveal the spectroscopic signature of the Frenkel-like S$_1$ exciton together with the residual intensity of the CT$_1$ exciton on its high energy side. In contrast, the other spectra show a broad spectral feature at $E - E_{\unit{HOMO}} = 2.45 \unit{eV}$. This is the signature of the CT$_1$ exciton before exciton dissociation.}
  \label{fgr:S5}
\end{figure*}

\newpage

\section{Supplementary Methods}
\subsection{Experimental Setup}
The Sc$_3$N@C$_{80}$ molecules were evaporated \textit{in situ}, using a Knudsen cell at a temperature of $520^{\circ}$C, onto an Ag(111) single crystalline surface prepared by cycles of Ar$^+$-sputtering and followed by annealing under UHV ($<10^{-9} \unit{mbar}$) conditions. The Sc$_3$N@C$_{80}$ multilayer film was prepared in several subsequent steps. First, a multilayer has been evaporated, which then has been annealed to a monolayer film by thermal desorption. At this preparation stage, the characteristic LEED pattern of the monolayer film can be observed \cite{Seidel2018}. At higher scattering energies, the substrate spots of the Ag(111) crystal can be observed again. Second, we evaporated a multilayer of Sc$_3$N@C$_{80}$ molecules, showing no additional diffraction spots in LEED and no valence bands of the substrate in photoemission.
The spectral shape of the Sc$_3$N@C$_{80}$ film is in very good agreement with literature \cite{Alvarez2002,Seidel2018}. For the K intercalation, a Saes Getters alkali metal dispenser was used. The effective potassium (K) concentration was estimated by comparison of the valence band UPS spectra with reference spectra from literature \cite{Alvarez2002}.
The time-resolved photoemission spectroscopy (trPES) measurements have been performed in a pump-probe setup with our fs-XUV light source \cite{Eich2014}, emitting p-polarized light pulses with $20 \pm 16 \unit{fs}$ pulse duration and $22.2 \unit{eV}$ photon energy. A time-delayed, sub-$50 \unit{fs}$, p-polarized, $3.2 \unit{eV}$ light pulse has been used to excite the electron system of the molecular film. This technique allows one to simultaneously observe both the dynamics of directly excited electrons (exciton dynamics) and the ensemble dynamics of the surrounding film (polaron dynamics), revealing the spectroscopic signature for the charge state of the optically excited excitons \cite{Stadtmuller2019}.
For both trPES and UPS measurements, our state-of-the-art hemispherical electron analyzer Specs Phoibos 150 has been used in normal emission.
Together with the chosen lens settings, this enables us to resolve a momentum regime of $k_{||} = \pm 0.5 ${\AA}$^{-1}$ around the $\bar{\Gamma}$-point. The molecular valence band spectra discussed in this paper are averaged over this momentum range.

\subsection{Details on the spectral analysis of excited states}
To extract the transient population dynamics of the different excited states from the acquired raw data, we generated energy distribution curves (EDCs) for each pump-probe time delay by integrating over the electron emission angle degree of freedom recorded by our hemispherical analyzer. Thereafter, the obtained 1D spectra were averaged over the multiple delay scans recorded within the measurement. In a next step, a Gaussian background was subtracted for every time delay of the scan. The data obtained this way is plotted in Fig.~\ref{fgr:S1} and \ref{fgr:S2} of the supplementary information (SI) for distinct time steps (dotted lines). Furthermore, by application of a sliding average for the delay time and a smoothing in energy, the overview graphs in figure 1a and 2a of the main text have been created from this data.
Subsequently, we analyzed the spectral shape of the exited states of the unsmoothed, background subtracted data by fitting each spectrum individually with three Gaussians. During the fitting procedure, the width and the peak positions were constrained to constant values. The peak positions and the peak width were optimized by iteratively repeating the fitting procedure with different, but time-independent width and position values. The only free fitting parameters are thus the peak heights of the three Gaussians. In Fig.~\ref{fgr:S1} and \ref{fgr:S2} of the SI, the found positions and widths are plotted together with the fit curve and the raw data for distinct time delays. The delay time dependent Gaussian peak area, being the observable describing the transient population of the respective excited states is shown in figures 1c and 2c of the main text.

\newpage

\subsection{Details on the analysis of the population decay of the excited states}
From the transient peak areas obtained by the orbital fit of the excited states presented in the previous section, population decay constants of the excited states CT$_2$ and CT$_1$ have been extracted by fitting their time dependent intensity evolution by the exponential fitting function $F_{\unit{CT}_2}(t)$ and $F_{\unit{CT}_1}(t)$:
\begin{align}
  F_{\unit{CT}_2}(t) = G(t) \otimes
  \begin{cases}
    I_0  &  t < t_0 \\
    I_{\unit{CT}_2} e^{-(t-t_0)/\tau_{\unit{CT}_2}}  &  t \geq t_0
  \end{cases}
\end{align}
\begin{align}
  F_{\unit{CT}_1}(t) = G(t) \otimes
  \begin{cases}
    I_0  &  t < t_0 \\
    I_{\unit{CT}_1} e^{-(t-t_0)/\tau_{\unit{CT}_1}} ( 1 - e^{-(t-t_0)/\tau_{\unit{CT}_1,\text{R}}})  &  t \geq t_0.
  \end{cases}
\end{align}
I$_0$ is the background intensity, $\tau_{\unit{CT}_2}$ the decay constant of the CT$_2$ level, $\tau_{\unit{CT}_1,\text{R}}$ the rise time of the intensity of the CT$_1$ level, and  $\tau_{\unit{CT}_1}$  the decay constant of the CT$_1$ level.
For the data analysis, the fit functions were convoluted with a normalized Gaussian function $G(t)$ with  $T_{\text{p}-\text{p}} = 70 \unit{fs}$. The latter value corresponds to the pump-probe cross correlation on the sample surface and was determined experimentally.
The best fitting results were obtained for  $\tau_{\unit{CT}_1,\text{R}} = \tau_{\unit{CT}_2}$. The corresponding fitting functions are shown as solid lines in figure 1c and 2c of the main manuscript. To describe the transient CT$_1$ population, a double-exponential decay with refilling has been assumed, similar to the method presented in the next section.

\subsection{Details on the analysis of the transient width dynamics of the occupied valence band states}
After having obtained the valence orbital widths as described in the caption of figure \ref{fgr:S3} of the SI, the decay constants of the fast decay channel $\tau_{\unit{TB}_2}$ and the slow decay channel $\tau_{\unit{TB}_1}$ can be extracted from the width $\Delta E$ using the following relation:
\begin{align}
  \Delta E(t) = G(t) \otimes
  \begin{cases}
    \Delta E_0  &  t < t_0 \\
    \Delta E_0 + \Delta E_{\unit{TB}_2} e^{-(t-t_0)/\tau_{\unit{TB}_2}} + \Delta E_{\unit{TB}_1} e^{-(t-t_0)/\tau_{\unit{TB}_1}} ( 1 - e^{-(t-t_0)/\tau_{\unit{CT}_2}})  &  t \geq t_0.
  \end{cases}
\end{align}
Here, $G(t)$ is a Gaussian function with a temporal width $T_{\text{p}-\text{p}} = 70 \unit{fs}$, introduced to consider the temporal broadening introduced by the pump and probe pulse. The fit function describes a double-exponential decay process with refilling to model the two subsequent decay steps observed in the excited states. The fit functions obtained using this method are plotted in figures 1e, 2e and 3c of the main text together with the respective data.

%\bibliography{Bibliography/C80_lib}% Produces the bibliography via BibTeX.